\documentclass[11pt]{article}

\usepackage{newtxmath}
\usepackage{bera}
\usepackage[T1]{fontenc}
\usepackage[utf8]{inputenc}

\usepackage{import}
\usepackage{url}
\usepackage[margin=1in]{geometry}
\usepackage{xcolor}
\usepackage{xspace}

\definecolor{darkblue}{rgb}{0.0,0.0,0.75}
\definecolor{lightgray}{rgb}{0.95,0.95,0.95}
\usepackage[bookmarks=true,bookmarksnumbered=false,bookmarksopen=true,hidelinks=false,colorlinks=true,citecolor=darkblue,linkcolor=darkblue]{hyperref}

\usepackage{paralist}
\usepackage{rotating}
\usepackage{graphicx}
\usepackage{bookmark}
\usepackage[sort]{cite}
\usepackage{epstopdf}
\usepackage{mathtools}
\usepackage{enumitem}
\usepackage{algorithmic}
\usepackage[operators, sets, adversary] {cryptocode}
\usepackage{framed}
\usepackage{float}
\usepackage{booktabs}
\usepackage{color,colortbl}
\usepackage{amsfonts}
\usepackage{relsize}
\usepackage{subfigure}
\usepackage[affil-it]{authblk}
\usepackage{cleveref}

\makeatletter
\renewcommand\AB@affilsepx{, \protect\Affilfont}
\makeatother

\usepackage[framemethod=tikz]{mdframed}
\usetikzlibrary{shadows}
\newmdenv[
	roundcorner=2pt,
	outerlinewidth=0.25pt,
	linecolor=black,
	shadow=true,
	shadowsize=4pt,
	frametitlerule=true,
	frametitlerulewidth=0.25pt,
	frametitlebackgroundcolor=gray!17,
	frametitlealignment={\vspace{0.006\linewidth}\center},
	font=\fontsize{9.75}{11.5}\selectfont,
	nobreak=true,
]{shadowbox}

\setlist[description]{listparindent=\parindent,leftmargin=0em,itemsep=0.3em,topsep=0.3em}

\newcommand{\formatNode}[1]		{\ensuremath{\mathsf{#1}}\xspace}

\newcommand{\A}					{\formatNode{A}}		% Alice
\newcommand{\B}					{\formatNode{B}}		% Bob

\ifxetex
	\renewcommand{\C}			{\formatNode{C}}		% Carole
\else
	\newcommand{\C}				{\formatNode{C}}		% Carole
\fi

\renewcommand{\S}				{\formatNode{S}}		% Server
\newcommand{\R}					{\formatNode{R}}		% Relay
\newcommand{\F}					{\formatNode{F}}		% Full node
\newcommand{\mint}				{\formatNode{unlock}}	% Mint
\newcommand{\burn}				{\formatNode{burn}}		% Burn
\newcommand{\sync}				{\formatNode{sync}}		% Sync
\newcommand{\tx}				{\formatNode{tx}}		% Transaction
\newcommand{\cp}				{\formatNode{cp}}		% Checkpoint
\newcommand{\addr}				{\formatNode{addr}}		% Address
\newcommand{\nullAddr}			{\texttt{null}}
\renewcommand{\H}				{\formatNode{H}}
\newcommand{\merkleVerify}		{\formatNode{MerkleVerify}}
\newcommand{\amount}			{\formatNode{amount}}
\newcommand{\sender}			{\formatNode{sender}}
\newcommand{\receiver}			{\formatNode{receiver}}
\newcommand{\mr}				{\ensuremath{\formatNode{tmr}}}		% TX Merkle root
\newcommand{\mmr}				{\ensuremath{\formatNode{cmr}}}		% chain/checkpoint MMR root
\newcommand{\quorumVerify}		{\ensuremath{\formatNode{QuorumVerify}}}
\newcommand{\sig}				{\ensuremath{\formatNode{sig}}}
\newcommand{\pks}				{\ensuremath{\formatNode{pks}}}

\newcommand{\formatMsg}[1]		{\ensuremath{\mathtt{#1}}\xspace}

\newcommand{\pobreqMsg}				{\formatMsg{PoB\text{-}Request}}	

\newcommand{\pobMsg}				{\formatMsg{PoB}}

\newcommand{\authnote}[2]{\textcolor{blue}{[#1: #2]}}

\newcommand{\mahdi}[1]{\authnote{Mahdi}{#1}}

\usepackage{fancyhdr}
\fancypagestyle{myfancy}{
	\lhead{}
	\rhead{}
	\chead{Horizon: A Trustless Bridge for Cross-Chain Transactions}
}
\renewcommand{\authnote}[2]{}

\begin{document}
	\sloppy
	
	\thispagestyle{empty}
	
	\title{\vspace{-1em}Horizon: A Gas-Efficient, Trustless Bridge for \\ Cross-Chain Transactions}

	\author[1]{\small Rongjian Lan}
	\author[1]{\small Ganesha Upadhyaya}
	\author[1]{\small Stephen Tse}
	\author[2]{\small Mahdi Zamani\thanks{This work was done when the author was affiliated with Yale University.}}
	\affil[1]{Harmony}
	\affil[2]{Visa Research}
	
	\date{}
	\maketitle
	
	\begin{abstract}
	With the rise of digital currency systems that rely on blockchain to ensure ledger security, the ability to perform cross-chain transactions is becoming a crucial interoperability requirement. 
	Such transactions allow not only funds to be transferred from one blockchain to another (as done in atomic swaps), but also a blockchain to verify the inclusion of any event on another blockchain. Cross-chain bridges are protocols that allow on-chain exchange of cryptocurrencies, on-chain transfer of assets to sidechains, and cross-shard verification of events in sharded blockchains, many of which rely on Byzantine fault tolerance (BFT) for scalability. Unfortunately, existing bridge protocols that can transfer funds from a  BFT blockchain incur significant computation overhead on the destination blockchain, resulting in a high gas cost for smart contract verification of events. 
	In this paper, we propose \emph{Horizon}, a gas-efficient, cross-chain bridge protocol to transfer assets from a BFT blockchain to another blockchain (e.g., Ethereum) that supports basic smart contract execution.	
\end{abstract}

	\vspace{1em}
	{\small \tableofcontents}
	
	\section{Introduction} \label{sec:intro}

A unifying vision of the distributed ledger technology is to connect multiple distributed systems together. While the disruption of financial infrastructure is still early, now is the time to bring many protocols together to scale their innovations for broader adoption through \emph{cross-chain transactions}. In particular, lending Bitcoin~\cite{nakamoto08bitcoin} for high-yield financial instruments on Ethereum~\cite{ethereum} has surpassed US \$1B on-chain management~\cite{Tokenize63:online}. Major chains like Cosmos, Polkadot and Cardano -- with promises for interoperability of contracts and assets -- recently launched the first phase of their mainnets~\cite{cosmos:whitepaper,CardanoH8:online,polkadot:whitepaper}. Cross-chain transactions are more than asset transfers via atomic swaps using hash time locked contracts (HTLCs)~\cite{HashTime62:online}; cross-chain transaction provide broader access of users and assets to any on-chain event happening on different distributed/decentralized ledgers. Such transactions happen frequently when exchanging cryptocurrencies~\cite{herlihy2018atomic,btcrelay:online}, transferring assets to sidechains~\cite{back2016transferring,nipopow,cryptoeprint:2018:1048}, or sharding blockchains~\cite{rapidchain:2018,omniledger:2018}.

Cross-chain transactions are typically facilitated by a \emph{chain relay} (or simply a \emph{relay}), an untrusted but highly-available node that relays block headers between two blockchains~\cite{btcrelay,near:bridge:online,xclaim:2019}. Namely, block headers of blockchain $A$ are constantly submitted to a smart contract on blockchain $B$ which implements a light client logic to verify the validity of the headers. Analogously, headers from blockchain $B$ are submitted to and verified by a smart contract on blockchain $A$.
An example is BTC Relay~\cite{btcrelay} which implements a uni-directional relay from Bitcoin to Ethereum. There, the Ethereum smart contract computes the difficulty of the submitted Bitcoin headers. A proof for the validity of each header amounts to checking it resides on the longest chain of submitted headers.

Light clients~\cite{nakamoto08bitcoin,nipopow,flyclient:2020} allow cross-chain transactions to happen efficiently. Such clients enable a user to generate short cryptographic inclusion proofs about past events recorded on a blockchain. %In a proof-of-work blockchain such as Bitcoin~\cite{nakamoto08bitcoin} and Ethereum~\cite{ethereum}, this can be done using a \emph{simplified payment verification (SPV)} client originally described in the Bitcoin paper~\cite{nakamoto08bitcoin}. An SPV proof consists of the sequence of all block headers which is sufficient to convince any verifier that the event is recorded on the honest chain, i.e., the longest, or more precisely, the most computationally-difficult chain. 
An important characteristic of most light client proofs that makes them suitable for cross-chain transactions is \emph{portability}, i.e., they can be forwarded to other nodes on other blockchain networks to convince them that a certain event was recorded on the source blockchain. 

While light clients for proof-of-work chains (such as Bitcoin~\cite{nakamoto08bitcoin} and Ethereum~\cite{ethereum}) have been proposed starting with Bitcoin's SPV client~\cite{nakamoto08bitcoin} and later improved in PoPoW~\cite{popow,nipopow} and FlyClient~\cite{flyclient:2020} for scalability, no light client has been proposed for chains that rely on Byzantine fault tolerance (BFT) for consensus such as proof-of-stake chains (e.g.,~\cite{cosmos:whitepaper,CardanoH8:online,polkadot:whitepaper,harmony:whitepaper,algorand17}). The proofs generated by such a client should be small enough to be efficiently verified by the smart contract on the destination blockchain. Otherwise, the gas cost associated with the proof verification could be high.

Making BFT light client proofs gas efficient is challenging. In most BFT chains, every block contains an aggregate signature (known as a \emph{quorum signature}) generated by a committee of validators who accepted that block at the time of consensus. While the aggregate signature is usually constant size and small, the verifier still requires the linear list of all signers' public keys (aka, \emph{quorum public keys}) to validate the signature. Since most BFT chains change the committee members periodically to ensure honest majority in the committee, the light client proof requires to include the linear list of all block signers in every proof, rendering the proof verification gas inefficient for smart contract on the destination blockchain.

%due to the following reasons: (1) While aggregate signatures contain a sufficient number of attestations, we do not know if those attestations actually come from validators that actually have stake. The only way to know this is if we know the balance (i.e., state) of every validator which is what a full node does.

\subsection{Our Contributions}

In this paper, we propose Horizon, a gas-efficient, cross-chain bridge protocol to transfer assets from a BFT blockchain to another blockchain (e.g., Ethereum) which supports basic smart contract execution. To achieve this, we make the following contributions:

\begin{itemize}
	\item We construct \textbf{a super-light client for BFT chains} that allows a client to prove to any external entity that a transaction has been recorded on the BFT chain by providing a cryptographic proof that is constant size in the length of the chain.
	
	\item We construct \textbf{a bridge smart contract} on the destination chain for atomic verification of super-light client proofs that guarantee a certain amount of tokens are locked on the BFT chain. The contract also can unlock/mint an equal amount of tokens on the destination blockchain once the verification succeeds.
		
	\item We construct \textbf{a relay node} which periodically transmits to the contract constant-size, checkpoint information as commitments to the BFT chain. This information allows the contract to later verify super-light client proofs submitted by the client to the contract. While the total amount of information submitted by the relay to the contract for all checkpoints is linear to the chain length, the frequency of checkpoints could be adjusted in practice to curb this overhead. 
		
	\item We propose \textbf{an efficient chain commitment} mechanism that allows the client to prove inclusion of a block in a blockchain with a constant-size commitment and logarithmic blockchain inclusion proofs.
	
	\item We further propose a \textbf{stateless bridge contract} design that allows the client to send a small, self-sufficient cross-chain transaction to the contract that does not require any pre-relayed checkpoint information. Our solution requires the client to include only a logarithmic-size (in the chain length) inclusion proof in its message, making it the first BFT bridge protocol that requires logarithmic-size, cross-chain proofs. 
\end{itemize}

\subsubsection{On PoW-to-BFT Transfers} 
Horizon further allows a client to transfer assets from a PoW chain (such as Bitcoin or Ethereum) to a BFT chain using FlyClient~\cite{flyclient:2020} logarithmic-size proofs. This, however, requires certain chain commitments (in the form of Merkle roots) already being included in every block header, which unfortunately, is not possible until a soft fork on Bitcoin and Ethereum includes these commitments in all future block headers. Until then, our bridge protocol adopts the SPV approach of Rainbow bridge~\cite{near:bridge:online}, where the relay node periodically sends all recent Bitcoin/Ethereum block headers to the smart contract on the BFT chain. While this incurs a significantly higher storage and computation overhead on the contract, we expect the significantly lower gas cost of most BFT chain (such as Harmony~\cite{harmony:whitepaper} and NEAR~\cite{near:bridge:online}) could justify such overhead until chain commitments become available on Bitcoin and Ethereum. %Moreover, one may \mahdi{The 7-day delay could even make it more efficient}. \mahdi{Do we know the precise cost of storing the last 7 days in Harmony contract?}
	\subsection{Our Model}

\paragraph{System Model.}
Consider a blockchain protocol $\A$, where a group of \emph{validators} agree on a chain via a BFT blockchain protocol such as~\cite{pass:hybridconsensus:2016,algorand17,solida:2017:opodis}, where validators participate in a BFT consensus protocol to agree on each block of transactions. In a permissionless setting, such a protocol typically proceeds in \emph{epochs}, where in each epoch, one or more BFT executions are followed by a reconfiguration step to randomly select a new group of validators to drive the next epoch. To obtain voting power in the BFT protocol, the reconfiguration protocol may establish identities through any Sybil-resistance mechanism such as PoW (e.g., in~\cite{pass:hybridconsensus:2016}) or proof-of-stake (PoS) (e.g., in~\cite{harmony:whitepaper}).

Let $\B$ denote another blockchain network that grows a valid chain based on any consensus mechanism (e.g., Nakamoto~\cite{nakamoto08bitcoin} or BFT) and provides basic smart contract support to execute arbitrary programs. We say a block $B$ of transactions added to $\B$'s chain is final when $B$ is permanently recorded on the chain. For example, in Bitcoin and Ethereum, this means that a sufficient number of blocks are appended to the chain after $B$, formalized as the most difficult chain principle by Garay~et al.~\cite{garay2017bitcoin}.

Our bridge system consists of (1) a \emph{client} $\C$ who can submit cross-chain transactions to $\A$ and $\B$, (2) a \emph{relay} $\R$ which periodically submits information about $\A$'s chain to $\B$, and (3) a \emph{full node} $\F$ that maintains an up-to-date copy of $\A$'s chain at any time and responds to queries from $\C$ and $\R$ about the chain.

\paragraph{Threat Model.} We consider a \emph{rational adversary} who may maliciously corrupt any party (i.e., $\C$, $\R$, and/or $\F$) or any coalition of them to prevent our protocol from achieving its goals, only if it benefits them. In other words, corrupt parties may deviate from the protocol in any arbitrary manner, e.g., by sending invalid or inconsistent messages or by intentionally remaining silent as long as the adversary gains some financial advantage by doing so. In addition, any of the parties may rarely crash (aka, fail stop) without the adversary's intervention, e.g., due to hardware/network failures. We assume that a pool of relays and full nodes is available to our protocol to replace $\R$ and $\F$, if they are detected by the protocol to be faulty.
We further assume that, at any time, both blockchains $\A,\B$ maintain standard blockchain immutability and double spending free properties, and guarantee that all valid transactions submitted to them will eventually be committed (i.e., written on the corresponding chain).  %In PoW chains, for example, this requires that, at any time, less than $1/2$ of the computational mining power (aka, hash rate) be controlled by the adversary. To ensure security against selfish mining~\cite{eyal2014}, one may alternatively assume less than $1/4$ adversarial computational power. Alternatively, a PoS chain may require less than $1/3$ of the total stake associated with validators in the network be controlled by the adversary.

\paragraph{Problem Definition.}
Client $\C$ wants to perform a cross-chain transaction $X_{\A\to\B}$ to transfer an amount of $x$ tokens from $\A$ to $\B$. The transaction consists of two on-chain transactions $T_\burn$ and $T_\mint$ to be recorded on $\A$ and $\B$ respectively.
A cross-chain bridge protocol is secure if and only if it guarantees the following property:

\begin{quote}
	\textit{Atomicity:} Transaction $X_{\A\to\B}$ is either committed or aborted. We say $X_{\A\to\B}$ is committed if and only if $T_\burn$ is committed (i.e., permanently recorded) on $\A$'s chain and $T_\mint$ is committed on $\B$'s chain.
	We say $X_{\A\to\B}$ is aborted if and only if $T_\burn$ and $T_\mint$ are both aborted (i.e., not committed on the corresponding chains).	
\end{quote}

\subsubsection{On The Feasibility of Trustless Bridges}
It has been shown that atomic cross-chain communication is impossible without a trusted third party (TTP)~\cite{cryptoeprint:2019:1128}. Intuitively, this is because the entity submitting $T_\mint$ to blockchain $\B$ after $T_\burn$ is recorded on blockchain $\A$ may fail to do so due to node crash or corruption, resulting in a permanent loss of funds. Since our protocol requires the client (i.e., the asset owner) to drive the cross-chain transaction, the rational adversary is not incentivized to prevent the client from submitting $T_\mint$ to $\A$. To protect the client's interest against rare crash failures, we assume that her assets on $\A$ (which she attempts to move to $\B$) are financially insured by a TTP (e.g., an insurance provider) who can verify the client's loss claims by accessing both blockchains. While such a TTP is theoretically required to solve the cross-chain bridge problem in general, our model removes the trust assumption from the critical cross-chain transaction loop, a suitable feature for practical settings.

	\section{Background and Related Work} \label{sec:background}

\subsection{Light Clients}
To verify that a blockchain is valid without participating in the mining process, a client may choose to download blocks from a miner or a \emph{full node} who holds a copy of the entire chain.
Currently, downloading and verifying all blocks in Bitcoin or Ethereum requires a node to download more than 200~GB of data, taking hours to synchronize the node's local blockchain~\cite{sync93:online}. Such a requirement causes long delays for regular clients and makes it nearly impossible for storage-limited clients to quickly verify transactions. 

The original Bitcoin design~\cite{nakamoto08bitcoin} describes a faster synchronization mechanism, known as \emph{simplified payment verification}, that allows efficient verification of transactions on the blockchain.
In Bitcoin~\cite{nakamoto08bitcoin} and Ethereum~\cite{ethereum}, block headers contain enough information to ensure that (1) the PoW is valid, (2) the block includes a certain transaction, and (3) the block is at a certain position on the correct chain. The transaction validation process utilizes a Merkle tree commitment to all transactions in a block which is stored in each block header. A light client does not verify all transactions in the entire chain and essentially relies on the assumption that the chain with the most proof-of-work contains only valid transactions and follows the rules of the system.

Kiayias~et al.~\cite{popow,nipopow} propose an interactive proof mechanism, known as proofs of proof-of-work (PoPoW) that allows a prover to convince a verifier with high probability in logarithmic time and communication that a chain contains a sufficient amount of work. The PoPoW protocol suffers from multiple drawbacks described by B\"unz~et al.~\cite{flyclient:2020} who propose a new solution, known as FlyClient, that overcomes the limitations of PoPoW. 

FlyClient uses a probabilistic sampling technique to randomly sample a logarithmic number of block headers (in the chain length) from a PoW-based blockchain with variable block difficulty. FlyClient uses an efficiently-updatable commitment mechanism, known as \emph{Merkle mountain range~(MMR)}~\cite{mmr}, that allows provers to commit to an entire blockchain with a small (constant-size) commitment while offering logarithmic block inclusion proofs with position binding guarantees.

\subsection{Cross-Chain Bridges}

\paragraph{Rainbow Bridge~\cite{near:bridge:online}.} NEAR's Rainbow bridge uses light clients to transfer ERC-20 tokens from Ethereum to NEAR's PoS blockchain and vice versa. For each chain, the Rainbow protocol deploys a smart contract that implements a light client and relay nodes that regularly sends block headers to the light client. Namely, Ethereum relays (aka, ETH-2-NEAR relays) sends every single Ethereum header to the NEAR contract, while the NEAR relays (aka, NEAR-2-ETH relays) sends one header every 4 hours to the Ethereum contract. As a result, both contracts can independently verify the inclusion of any event on the other chain. 

While the Rainbow bridge allows trustless cross-chain transfers, it has multiple performance and security drawbacks. First, the relays need to constantly forward Ethereum and NEAR headers to the smart contracts and this comes at a large gas cost on both chains. Second, to avoid an ever-growing state of the NEAR contract, the bridge limits the number of synced headers to only seven days. If the contract also limits the verification to only the seven-day log, then this significantly lowers the security of the light client as a malicious Ethereum relay can now create a fake chain of only seven days worth of headers appended to a valid prefix and present it to the NEAR contract. If the contract also incorporates previous seven-day logs from old blocks on the NEAR blockchain, then the large gas cost overhead would again come into play. Third, Rainbow's Ethereum contract does not verify all validator signatures on every NEAR header submitted to the contract.

\paragraph{XCLAIM~\cite{xclaim:2019}.} XCLAIM is a framework for trustless, cross-chain exchanges, where a smart contract on each chain governs the exchanges between the two chains (e.g., Bitcoin and Ethereum) and punishes malicious parties by seizing their collateral in favor of honest parties. The XCLAIM model consists of three main entities: A client who wishes to move funds from Bitcoin to Ethereum, a vault that locks the Bitcoin funds received from the client, and an Ethereum relay contract known as BTCRelay~\cite{btcrelay} which stores Bitcoin block headers to allow verification of SPV proofs. 

The protocol starts with the vault locking up sufficient collateral on the Ethereum smart contract. The client then sends her Bitcoins to the vault and submits a proof to the contract showing that the transaction has been recorded on the Bitcoin blockchain. The chain relay verifies this proof and confirms to the contract that the lock has been executed correctly. Finally, the contract releases Ethereum tokens to Alice.

\paragraph{tBTC~\cite{tBTC}.} tBTC is a multi-wallet, multi-signer protocol that provides a Bitcoin-backed bearer asset on Ethereum. %tBTC is the only link between the Bitcoin chain and the host chain which runs as a set of smart contracts on the host chain. 
tBTC attempts to remove single points of failure by geographically distributing signers and aiming for a multi-wallet scheme. 
The signers use a multi-party threshold ECDSA protocol to collectively create a signing group wallet which is created by randomly selecting a set of signers from the eligible pool of signers. tBTC relies on collateral to prevent signers from deviating the protocol as in the following cases:
(1) To liquidate deposits in case they are in danger of undercollateralization;
(2) To punish a signing group if it signs an unauthorized piece of data is once distributed key generation is complete;
(3) To punish a signing group that fails to produce a signature for the system when requested; and
(4) To ensure a depositor is refunded if the signing group fails to form.
As a result, tBTC requires a complicated mechanism for detecting and dealing with undercollateralized signers.

\paragraph{Waterloo~\cite{waterloo:bridge}.} Kyber's Waterloo is a cross-chain bridge between Ethereum and EOS~\cite{eos:whitepaper}. An Ethereum smart contract serves as a light client that only verifies EOS block headers. The consensus protocol of EOS is based on the delegated PoS mechanism~\cite{delegated:pos}, where EOS token holders continuously vote (i.e., delegate) for their favorite block producers. Instead of relaying all EOS block headers, the Waterloo relay only relays the changes in the set of EOS block producers.

\subsection{Harmony Blockchain}

We build our solution for transferring assets from a BFT blockchain to a PoW blockchain in the context of the Harmony blockchain~\cite{harmony:whitepaper} as our BFT protocol and Ethereum as our PoW blockchain. Harmony is a sharded PoS blockchain protocol inspired by research results including but not limited to RapidChain~\cite{rapidchain:2018} and OmniLedger~\cite{omniledger:2018}. The protocol execution is divided into predetermined time intervals (e.g., 24 hours), known as \emph{epochs}. In each epoch, the network is partitioned into a set of \emph{shards} each of which maintains a separate blockchain in parallel to other shards. 

The validators of shards are randomly sampled using a distributed random generation (DRG) protocol executed at the end of every epoch by a special shard known as the \emph{beacon shard}, which itself gets re-elected in every epoch.
The beacon shard is also where the configuration of the network (i.e., validator identities and shard assignments) are stores. Moreover, the beacon shard is where token holders deposit their stakes to become validators in the consensus protocol. 

Each epoch consists of multiple executions of a BFT consensus protocol in each shard, where a block of transactions is generated and appended to the shard's chain after every execution. The set of validators and shards is fixed throughout each epoch but may change between epochs based on a shard reconfiguration protocol executed by the beacon shard at the end of each epoch. The last block of the beacon chain in every epoch is called an \emph{epoch block} which stores the identities (i.e., public keys) of all shard members for the next epoch. Each identity consists of the node's ECDSA address, its BLS public key, and its stake in the consensus, represented as a decimal number between 0 and 1.

For every BFT consensus execution, a leader is selected based on the randomness generated by the DRG at the end of the previous epoch.\footnote{Harmony's DRG uses a verifiable random function (VRF)~\cite{micali:1999:vrf} in conjunction with a verifiable delay function (VDF)~\cite{vdf:2018} construction to generate randomness with linear communication complexity and delay the revelation of the randomness generated by the VRF to prevent a malicious leader from biasing the randomness by cherry-picking a subset of the VRF random numbers initially generated by the validators~\cite{harmony:whitepaper}.}
In every run of the consensus protocol, the leader runs the aggregate BLS signature protocol of Boneh~et al.~\cite{boneh:2001:ssw:647097.717005,aggregate:bls:2003} to collect the validators’ votes in a constant-sized, threshold aggregate signature and then broadcast it to the shard. The aggregate signature is included in the block header for later block verification. After the new block is committed to the shard chain (i.e., the chain maintained by the shard), the shard validators send the block header to the beacon shard who verifies the header contents (i.e., the previous hash and the aggregate signature) and broadcast it to all shards to facilitate future cross-shard verification.

	\section{Our Solution}

Consider a client $\C$ who wants to perform a cross-chain transaction $X_{\A \to \B}$ to transfer an amount $x$ of her coins from blockchain $\A$ (i.e., the Harmony blockchain) to blockchain $\B$ (e.g., the Ethereum blockchain). Our bridge protocol, Horizon, consists of a smart contract $\S$ deployed on blockchain $\B$ as well as a relay node $\R$ who periodically syncs $\S$ with epoch block headers from $\A$. 
As shown in Figure~\ref{fig:protocol}, Horizon consists of two parts: (1) Relay/contract sync, and (2) Cross-chain transaction. In the following sections, we describe each part in detail.

\begin{figure}[t]
	\centering
	\includegraphics[width=0.85\columnwidth]{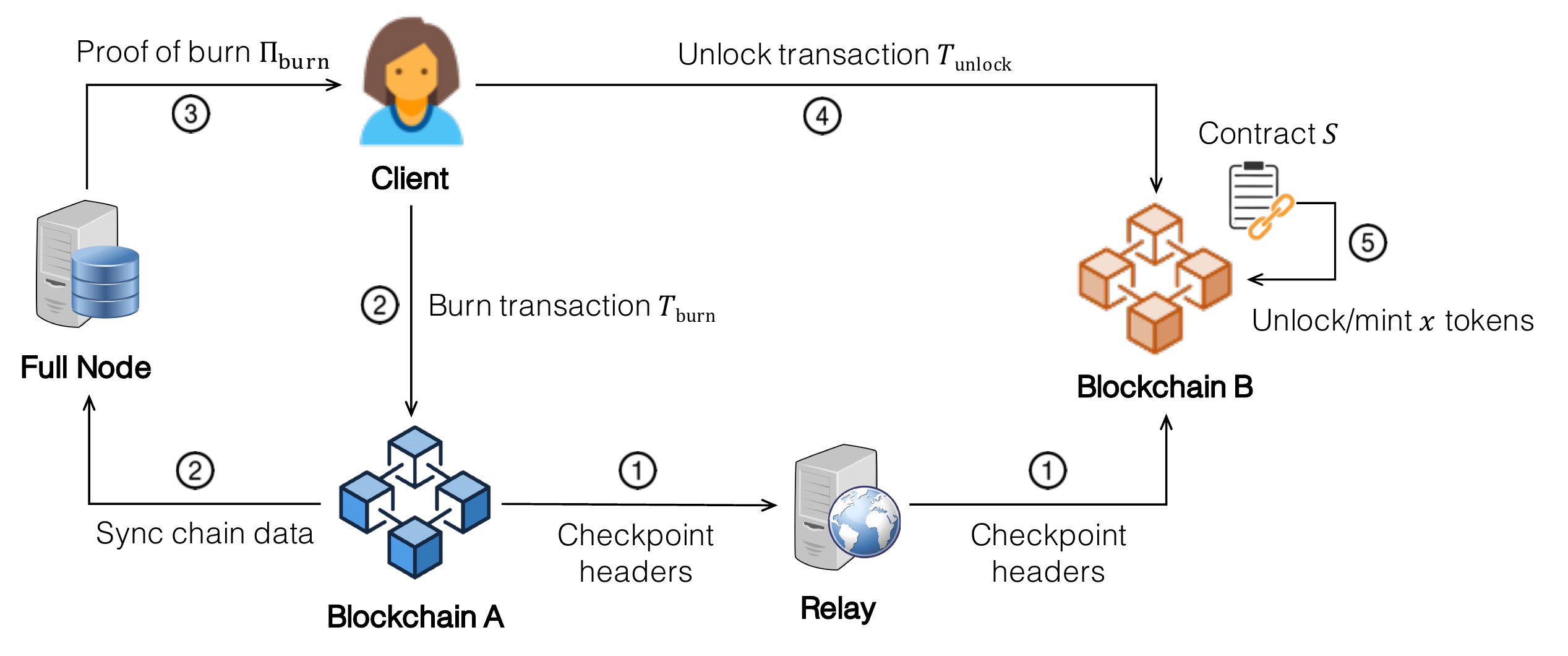}
	\caption{Horizon Cross-Chain Bridge Protocol}
\end{figure}

\begin{figure}
	\begin{shadowbox}[frametitle={\textsc{Horizon} Cross-Chain Bridge Protocol}]
		Client $\C$ performs a cross-chain transaction $X_{\A \to \B}$ to transfer an amount $x$ of her coins from blockchain $\A$ to blockchain $\B$. Smart contract $\S$ on $\B$ verifies and unlocks the transferred coins, and relay $\R$ periodically syncs $\S$ with checkpoint data from $\A$.
		
		\begin{itemize}[topsep=0.5em, leftmargin=0em, itemsep=1.25em]
			\item[] \textbf{\underline{Relay/Contract Sync}}\\[0.5em]
			Let $\addr_\R$ denote $\R$'s wallet address on $\B$. The following protocol is executed between $\R$ and $\S$:
			\begin{enumerate}		
				\item At every checkpoint block header $B_i$, relay $\R$ sends a sync transaction $T_\sync$ to $\S$, where 
				\vspace{-0.75em}
				\[T_\sync: (\addr_\R \to \nullAddr; B_i).\] 
				$B_i$ includes block height $i$, quorum signature $B_i.\sig$, quorum public keys $B_i.\pks$,  and Merkle root $B_i.\mmr$ created on all block header between $B_{i-\Delta}$ and $B_i$, where $\Delta$ is the block distance between two checkpoint blocks.
				
				\item Upon receiving $T_\sync$ from $\R$, contract $\S$ replaces $\R$ and aborts if any of the followings are true:
				\begin{enumerate}
					\item $i \neq j+1$, where $j$ is the height of the last checkpoint block received, or
					\item $\quorumVerify(B_i.\sig, B_i.\pks) = 0$.
				\end{enumerate}
				Otherwise, $\S$ stores $B_i$ in the contract's state.
			\end{enumerate}
			
			\item[] \textbf{\underline{Cross-Chain Transaction}}\\[0.5em]
			Let $\addr_\C$ denote client $\C$'s wallet address on both $\A$ and $\B$. The client performs the following protocol with full node $\F$ and smart contract $\S$: 
			
			\begin{enumerate}[topsep=0.25em, leftmargin=4em, itemsep=1em, label=Stage \Roman*:, ref=\Roman*]
				\item Burn
				\begin{enumerate}[leftmargin=-1em,itemsep=0.25em, label=\arabic*., ref=\theenumi{}--\arabic*]
					\item Client $\C$ submits a burn transaction $T_\burn: \addr_\C \xrightarrow{x} \nullAddr$ to $\A$ to burn $x$ of her coins on $\A$.
					
					\item Once $T_\burn$ is confirmed on $\A$, client $\C$ sends [$\pobreqMsg,h$] to $\F$, where $h = \H(T_\burn)$. \label{step:fullnodequery}
					
					\item Upon receiving $[\pobreqMsg, h]$ from $\C$, full node $\F$ performs the following steps:
					
					\begin{enumerate}[topsep=0.5em, itemsep=0.25em, label=(\alph*)]
						\item Find block header $B_k$ on $\A$ that includes $h$;
						
						\item Generate a Merkle proof $\Pi_B$ using the checkpoint Merkle root $B_i.\mmr$, where ${k \in (i-\Delta, i]}$; %showing that block $B$ was recorded on $\A$;
						
						\item Generate a Merkle proof $\Pi_T$ using the transaction Merkle root $B_k.\mr$;%showing that $T_\burn$ is included in $B_k$;
						
						\item Send $[\pobMsg, \Pi_\burn, i, B_k]$ to $\C$, where $\Pi_\burn = (\Pi_B, \Pi_T)$.
					\end{enumerate} 
					
					\item Upon receiving $[\pobMsg, \Pi_\burn, i, B_k]$ from $\F$, client $\C$ submits an unlock transaction $T_\mint$ to $\S$, where
					\vspace{-0.75em}	
					\[T_\mint: (\nullAddr \xrightarrow{x} \addr_\C; ~T_\burn; ~\Pi_\burn; ~i; ~B_k).\]
					
					\item If $T_\mint$ fails, then $\C$ picks a different full node and repeats from Step~\ref{step:fullnodequery}.
				\end{enumerate}			
				\item Unlock
				\begin{enumerate}[resume,topsep=0em,leftmargin=-1em, label=\arabic*.]
					\item Upon receiving $T_\mint: (\nullAddr \xrightarrow{x} \addr_\C; ~T_\burn; ~\Pi_\burn; ~i; ~B_k)$, contract $\S$ does the following steps:
					\begin{enumerate}[topsep=0.5em,itemsep=0.25em,label=(\alph*)]
						\item Parse $\Pi_\burn$ as $(\Pi_B, \Pi_T)$;
						\item Fetch $B_i$ from the state of contract $\S$;
						\item Abort if any of the following conditions are true: 
						\begin{itemize}[topsep=0.5em, itemsep=0.4em, label={--}]
							\item $T_\burn.\sender \neq T_\mint.\receiver$, ~$T_\burn.\amount \neq x$,
							\item $\merkleVerify(\Pi_B, B_k, B_i.\mmr) = 0$, where $B_i.\mmr$ is checkpoint Merkle root,
							\item $\merkleVerify(\Pi_T, T_\burn, B_k.\mr) = 0$, where $B_k.\mr$ is $B_k$'s transactions Merkle root.
							%\item $\quorumVerify(B_k.\sig, B_k.\pks) = 0$.
						\end{itemize}
						\item Unlock $x$ to $\addr_\C$'s balance.
					\end{enumerate}
				\end{enumerate}
			\end{enumerate}
			%			\item[] \textbf{\underline{PoB-Verify($\Pi_\burn, i$)}}\\[0.5em]
			%			Output 1 if and only if all of the following conditions are true, otherwise return 0:
			%			\begin{itemize}[topsep=0.5em, label={--}]
			%				\item $\merkleVerify(\Pi_B,M_i) = 1$,
			%				\item $\merkleVerify(\Pi_T,M_i) = 1$,
			%				\item $T_\burn.\sender = T_\mint.\receiver,$
			%				\item $T_\burn.\amount = x.$
			%			\end{itemize}
		\end{itemize}
	\end{shadowbox}
	
	\vspace{-1.25em}	
	\caption{Horizon Cross-Chain Bridge Protocol}	
	\label{fig:protocol}	
\end{figure}

\subsection{Relay/Contract Sync}
At the end of each epoch (i.e., every 24 hours), $\R$ sends to the contract the most recent epoch block header $B_i$ which is maintained by the beacon shard. This block contains sufficient information to allow the contract to later verify the inclusion of any transaction on $\A$. In our bridge protocol, the contract verifies the inclusion of a \emph{burn transaction}, denoted by $T_\burn$, submitted by the client. $T_\burn$ essentially transfers $x$ coins on $\A$ to a $\nullAddr$ address, i.e., deletes the coins permanently. We refer to an inclusion proof for a burn transaction as a \emph{proof of burn (PoB)}. 

For ease of presentation and without loss of generality, we consider only one shard (i.e., the beacon shard) and describe Horizon for the case when $T_\burn$ is recorded on the beacon chain.\footnote{In the Harmony blockchain, the beacon shard also grows a transaction chain similar to other shards.}

\subsubsection{Proof of Burn (PoB)}
An inclusion proof for $T_\burn$ should convince the verifier (i.e., the contract) that (1) The transaction is included in a block with header $B_k$, and (2) The block corresponding to $B_k$ is included on $\A$'s chain. 
The former can be proved using a transaction Merkle proof and verified using the Merkle root stored in $B_k$. The latter, however, requires a new chain commitment added to the epoch block $B_i$. Inspired by FlyClient~\cite{flyclient:2020}, we use a Merkle tree variant, known as a \emph{Merkle Mountain Range (MMR)}~\cite{mmr}, over all block headers added to the blockchain between two epoch blocks. This allows $\S$ to verify the inclusion of $B_k$ within the epoch using the root of the MMR stored at $B_i$. \mahdi{why an MMR and not a regular MT?}

\subsubsection{Checkpoint Blocks}
The confirmation latency of a cross-chain transaction depends mainly on the rate at which $\R$ submits epoch block headers to $\S$. Since one epoch block is created every 24 hours, then a cross-chain transaction would take about 12 hours to be confirmed in expectation. 
To reduce the confirmation latency, we propose to create periodic \emph{checkpoint blocks} in the middle of epochs on $\A$. A block is called a checkpoint block if its header contains an MMR root calculated over all block headers added to the chain since the previous checkpoint block. Therefore, epoch blocks are considered checkpoint blocks.

Every checkpoint block header $B_i$ includes the following fields the first four of which are included in all blocks:
\begin{enumerate}[itemsep=0em]
	\item Block height $i$,
	\item Quorum signature $B_i.\sig$, an aggregate BLS signature created by the consensus validators,
	\item Quorum public keys $B_i.\pks$, which lists the public key address of every consensus validator,
	\item Transaction Merkle root $B_i.\mr$ created on all transaction included in $B_i$,
	\item Checkpoint Merkle root $B_i.\mmr$ created on all block header between $B_{i-\Delta}$ and $B_i$, where $\Delta$ is the block distance between two checkpoint blocks.
\end{enumerate}

\subsubsection{Checkpoint Verification}
At every checkpoint block header $B_i$, the relay sends a sync transaction $T_\sync$ to $\S$ as defined in Figure~\ref{fig:protocol}. Upon receiving $T_\sync$ from $\R$, contract $\S$ verifies that the checkpoint information is valid, otherwise it replaces $\R$ with another relay and aborts. Namely, $\S$ does the replacement and aborts if any of the following conditions are true:
\begin{enumerate}
	\item $i \leq j$, where $j$ is the height of the last checkpoint block received,
	\item $\quorumVerify(B_i.\sig, B_i.\pks) = 0$.
\end{enumerate}
Otherwise, $\S$ stores $B_i$ in the contract's state for future cross-chain verification requests.

	%\mahdi{How/when/where is the aggregate BLS public key generated? The new shard validators should run DKG on the beacon shard, where the aggregate public key is created. The beacon shard then commits to this public key. The relay should forward this signature along with other information to the contract so the contract can verify the authenticity of the public key. }

\subsection{Cross-Chain Transactions} 
This part of the protocol is initiated by the client $\C$ submitting $T_\burn$ to blockchain $\A$. Once the transaction is confirmed on $\A$, $\C$ sends a request to a full node $\F$ who always maintains a full, up-to-date copy of $\A$'s chain. Upon receiving the request, $\F$ finds the block $B_k$ that includes $T_\burn$. Then, it generates two Merkle proofs $\Pi_B$ and $\Pi_T$ that together form the PoB, denoted by $\Pi_\burn$. The former is generated using the checkpoint Merkle root $\mmr_i$ stored at the first checkpoint block after $B_k$, denoted by $B_i$. The latter is generated using the transaction Merkle root $B_k.\mr$. Finally, $\F$ sends $\Pi_\burn$, $i$, and $B_k$ to $\C$ who creates an unlock transaction $T_\mint$ and submits it to blockchain $\B$. Since the full node is not trusted, this transaction may fail, in which case the client replaces $\F$ with another full node and repeats the process.

\subsubsection{PoB Verification and Asset Unlocking} 
Upon receiving $T_\mint$, contract $\S$ fetches $B_i$ from its state (checkpoint block height $i$ is specified by $\C$ in $T_\mint$) and verifies the validity of the burn transaction against the unlock transaction through the checks listed in Figure~\ref{fig:protocol}. If all checks are successful, then the contracts unlocks $x$ tokens on $\B$ (maintained on $\S$'s state) and delivers them to $\C$'s address on $\B$.
	
\subsection{Multi-Relay Model}
The single-relay bridge model exposes cross-chain transactions to denial-of-service scenarios which could make the bridge protocol completely non-functional. One may alternatively choose to employ a group of relay nodes which redundantly submit the same checkpoint information to the smart contract. Unfortunately, this significantly increases the gas cost of relaying information to the contract as it needs to compare/store an amount of information that grows linearly with the redundancy factor (i.e., the number of relays).

Instead of multiple relay nodes actively submitting checkpoint information to the contract at the same time, we propose to have only one node relay the information at any time and have the other $n-1$ relays read and verify the contract's state after every regular relay event. If the first relay fails to send proper checkpoint information, the second relay would take its role and so forth.   

\subsection{Stateless Bridge Contract} 
In some scenarios, one may prefer to implement a bridge protocol without a relay. For example, if the number of cross-chain transactions is small, it might not justify the overhead (i.e., the gas cost) of constantly sending checkpoint information to the smart contract. That's because the number of checkpoint blocks grows linearly with the chain length. 

On the other hand, due to the block gas limit, the client cannot include the list of checkpoint blocks for the entire chain in its unlock transaction $T_\mint$. While the client can split the transaction data into multiple transactions/blocks to avoid the gas limit issue, this strategy likely still results in an unreasonable gas cost for a single cross-chain transactions. Moreover, splitting a transaction across multiple blocks requires waiting for multiple block intervals, and thus significantly increasing the latency of the overall transaction.

To allow relay-free cross-chain transactions, we propose to add another layer of chain commitments to every checkpoint block in the form of an MMR constructed over all checkpoint blocks. This allows the prover (i.e., the client) to prove to the verifier (i.e., the contract) that a checkpoint block is included in a missing chain of checkpoint blocks. The client proof now consists of the following information:

\begin{enumerate}[itemsep=0em]
	\item The header of the block $B_\tx$ containing $T_\burn$.
	\item The header of the first checkpoint block $B_\cp$ after $B_\tx$.
	\item A Merkle proof showing that $T_\burn$ was included in $B_\tx$.
	\item A Merkle proof showing that $B_\cp$ was included in the chain of checkpoint blocks. 
\end{enumerate}

	\bibliographystyle{unsrt}
	\bibliography{main}

\end{document}